\documentclass[draftclsnofoot,11pt,onecolumn]{IEEEtran}
\usepackage{amssymb}
\usepackage[dvips]{graphicx}
\usepackage[cmex10]{amsmath}
\usepackage{amsthm}
\usepackage{latexsym,bm}
\usepackage{color}
\usepackage{subfigure}
\usepackage{multirow}
\usepackage{longtable}
 \def\bh{{\mathbf{h}}}

  \def\br{{\mathbf{r}}}

 \def\bn{{\mathbf{n}}}

\def\bB{{\mathbf{B}}} \def\bI{{\mathbf{I}}}

\theoremstyle{plain}

\theoremstyle{definition}\newtheorem{defi}{Definition}
\theoremstyle{definition}

\begin{document}
\title{Performance Analysis of Protograph-based\\LDPC Codes with Spatial Diversity}
\author{\IEEEauthorblockN{\fontsize{11pt}{\baselineskip}\selectfont
{Yi Fang$^1$,\ Pingping Chen$^1$,\ Lin Wang$^1$, \emph{Senior Member, IEEE},
\\ Francis C.M. Lau$^2$, \emph{Senior Member, IEEE},\ and Kai-Kit Wong$^3$, \emph{Senior Member, IEEE}}}
\IEEEauthorblockA{\normalsize{
\\$^1$College of Information Science and Technology, Xiamen University,
361005 Xiamen, P. R. China
\\$^2$Department of Electronic and Information Engineering, Hong Kong Polytechnic University, Kowloon, Hong Kong
\\$^3$Department of Electronic and Electrical Engineering, University College London, UK
}}}

\maketitle

\begin{abstract}
In wireless communications, spatial diversity techniques, such as space-time block code (STBC) and
single-input multiple-output (SIMO), are employed to strengthen the robustness of the transmitted signal against channel fading.
This paper studies the performance of protograph-based low-density parity-check (LDPC) codes with receive antenna diversity. We first propose a modified version of the protograph extrinsic information transfer (PEXIT) algorithm and use it
for deriving the threshold of the protograph codes in a single-input multiple-output (SIMO) system. We then calculate the decoding threshold and simulate the bit error rate (BER) of two protograph codes (accumulate-repeat-by-3-accumulate (AR3A) code and accumulate-repeat-by-4-jagged-accumulate (AR4JA) code), a regular $(3,6)$ LDPC code and two optimized irregular LDPC codes. The results reveal that the irregular codes achieve the best error performance in the low signal-to-noise-ratio (SNR) region and the AR3A code outperforms all other codes in the high-SNR region. Utilizing the theoretical analyses and the simulated results, we further discuss the effect of the diversity order on the performance of the protograph codes. Accordingly, the AR3A code stands out as a good candidate for wireless communication systems with multiple receive antennas.
\end{abstract}


\section{Introduction}
In wireless communications, fading is a major factor that deteriorates the quality of signal transmission. Many methods have been proposed to mitigate the effect of fading. Of particular interest are the multi-antenna technologies which  can provide high diversity gain and spatial multiplexing gain \cite{Telatar1999}.
Recently, a wealth of research has investigated the interplay between forward error correction (FEC) and spatial diversity. As a type of superior FEC code,
 low-density parity-check (LDPC) codes are known to perform near the Shannon limit over the additive white Gaussian noise (AWGN) channel \cite{905935}. However, LDPC codes that perform well in AWGN channels may not do so in a fading
environment \cite{1412036}. To overcome this weakness, LDPC codes have been studied in fading environments and capacity-approaching LDPC codes have also been designed for single-input multiple-output (SIMO) channels  and multiple-input multiple-output (MIMO) channels using density evolution \cite{4109664,4746596,5439356}
and  extrinsic information transfer (EXIT) function \cite{1291808,5407603}, respectively.

Many of the capacity-approaching LDPC codes, however, are irregular and hence suffer from two main
drawbacks --- high error floor and nonlinear encoding. Recently, some research has successfully reduced the error floor of short-block-length LDPC codes which can then attain outstanding performance down to an error rate of $10^{-5}$ \cite{4357432}.
Meanwhile, the quasi-cyclic LDPC codes that permit linear encoding have been proposed for MIMO channels \cite{4543051}.
In addition, a novel class of LDPC code, namely multi-edge type (MET) LDPC code, has been introduced  \cite{Member_multi-edgetype}. As one subclass of the MET-LDPC code, the protograph-based LDPC code has emerged as a promising FEC scheme due to its excellent error performance and low complexity \cite{Thorpe2003ldp}.  Two types of protograph codes, namely accumulate-repeat-accumulate (ARA) code and accumulate-repeat-by-4-jagged-accumulate (AR4JA) code, which can realize linear encoding and decoding have been proposed by Jet Propulsion Laboratory (JPL) \cite{1577834,4155107,5174517,5454136}. In \cite{4411526}, a protograph EXIT (PEXIT) algorithm has been introduced to predict the threshold of protograph codes over the AWGN channel. While the protograph codes have further been studied under Rayleigh fading channels \cite{6133952}, to the best of our knowledge, little is known about the analytical performance for protograph codes  under fading conditions and antenna diversity has never been considered.

In this paper, we aim to investigate the performance of the protograph codes over a SIMO Rayleigh fading
channel. To do so, we propose a modified PEXIT algorithm for analyzing the protograph LDPC code over a fading
environment. We then analyze the decoding thresholds of the accumulate-repeat-by-3-accumulate (AR3A) code, the AR4JA code, the regular $(3,6)$ LDPC code and two optimized irregular LDPC codes \cite{4109664} and use the thresholds to predict the error performance of the codes.
The results show that the irregular LDPC codes outperform the other codes in the low signal-to-noise-ratio (SNR) region. However, in the high-SNR region, the AR3A code possesses the best error performance. Besides, we also study the performance of the AR3A code and the AR4JA code with different diversity orders
based on (i) the mean and variance of the log-likelihood-ratio (LLR) values,
(ii) the PEXIT analysis, and
(iii) the bit-error-rate (BER) simulations.
We find that the additional gain becomes smaller as we increment the diversity order
and hence increase the system complexity.

We organize the remainder of this paper as follows.
In Section~\ref{sect:review}, we present the system model over the SIMO fading channels and in Section~\ref{sect:PEXIT}, we describe our modified PEXIT algorithm for analyzing protograph codes in the fading environments.
In Section~\ref{sect:analysis}, we analyze the decoding threshold and the initial LLR distribution of two conventional protograph codes.
We show the simulation results in Section~\ref{sect:sim_dis} and finally we give the concluding remarks in Section~\ref{sect:conclusion}.

\section{System Model}\label{sect:review}
\begin{figure}[h]
\center
\includegraphics[width=3.5in,height=1.3in]{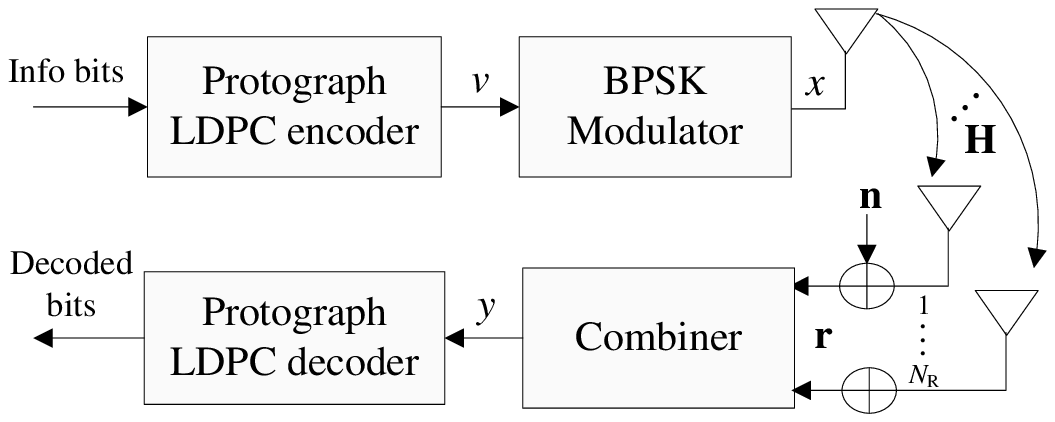}
\vspace{-0.3cm}
\caption{System model of the protograph codes over the SIMO Rayleigh fading channels.
}
\label{fig:Fig.1}
\end{figure}

The system model being considered in this paper is illustrated in Fig.~\ref{fig:Fig.1}. Referring to this figure, the information bits (Info bits) are firstly encoded by the (punctured) protograph LDPC code. Then the binary coded bits $v \in \{0,1\}$ are passed to a binary-phase-shift-keying (BPSK) modulator, the output of
which is given by
$x = (-1)^v \in \{+1, -1\}$. The modulated signal $x$ is further sent through a SIMO fading channel with one transmit antenna and $N_{\rm R}$ receive antennas.

 We denote $\bh$ as a channel realization vector of size $N_{\rm R} \times 1$, the entries of which are  complex independent Gaussian random variables with zero-mean and variance $1/2$, i.e., ${\cal N}(0,1/2)$, per dimension.
 {\color{red} Moreover, $\bh$ is assumed to be independent over time \cite{4109664,4746596,1291808,5407603}}.
We define  $\bn$ as a $N_{\rm R} \times 1$ complex AWGN vector with zero-mean   and covariance matrix
${\mathbb E}(\bn \bn^\dagger) = (N_0/2)\bI_{N_{\rm R}} = \sigma_n^2 \bI_{N_{\rm R}}$
where $\mathbb E (\cdot)$ is the expectation operator, $^\dagger$ denotes the transpose conjugate operator, and $\bI_{N_{\rm R}}$ represents
the $N_{\rm R} \times N_{\rm R}$ identity matrix. Then, the  $N_{\rm R} \times 1$ receive signal vector, denoted by $\br$, is given by
\begin{equation}\label{eq:receiver}
\br = \bh x + \bn.
\end{equation}
Note that up to now, the time index has been omitted for clarity.

At the receiver, we assume that the $N_{\rm R}$ received signals in $\br$
are combined using the maximum-ratio-combining (MRC) method \cite{875282} or the equal-gain-combining (EGC) method \cite{871398}{\color{red}\footnote{\color{red}
Minimum mean square error (MMSE) combiner is particularly useful in mitigating interference in the frequency selective channels. Thus, a MMSE combiner is suitable for the cases in which interference exists \cite{1424518}.
In this paper, however, interference does not exist in the channel.
Consequently, we apply the simpler
MRC and  EGC \cite{875282,871398}, which have also been used to
process the received signals over interference-free Rayleigh fading channels incorporated with multiple antennas and LDPC codes \cite{4109664,4746596}.
}.}
Afterwards, the combined signals $y$
 are sent to the LDPC decoder for finding the valid codewords.
 We do not consider interleaving in our system{\color{red}\footnote{\color{red} Note
 that in a practical environment, interleaving may be required to ensure that the channel gains
 for the code bits in the same codeword are independent.}}.
 Moreover, we assume that (i) the channels formed by different transmit-receive antenna pairs are independent and (ii) each receive antenna possesses perfect channel state information (CSI) which is changing sufficiently rapidly to satisfy ergodicity {\color{red}\cite{1291808}}. The capacity of an ergodic SIMO channel is further given by \cite{Telatar1999}
\begin{equation}\label{eq:Capcity}
C = {\mathbb E} \left(\log_2 \det \left(\bI_{N_{\rm R}} + \frac{E_s} {N_0} \bh \bh^\dagger\right)\right)~~ {\rm bits/s/Hz},
\end{equation}
where $E_s$ is the average energy per transmitted symbol and $\det(\cdot)$ is the determinant operator. This capacity can be evaluated by using Monte Carlo simulation, i.e., by generating a large number of independent channel realizations and computing their average capacity value. In practical applications, we should choose a modulation technique and a code rate $R$ such that the product of
the modulation order $M_c$ and the code rate equals the channel capacity, i.e., $ M_c R = C$. For BPSK modulation, $M_c=1$ and hence we have $R = C$.
Moreover, the definition of $E_b / N_0$ ($E_b$ is the average energy per information bit) used in this paper is the same as that used in \cite{1291808}, i.e.,\begin{equation}\label{eq:SNR}
\frac{E_b} {N_0} = \frac{(N_{\rm R}/R) E_s} {N_0}.
\end{equation}

\section{Modified PEXIT Algorithm for Protograph Codes}\label{sect:PEXIT}
The conventional EXIT function \cite{1291808} has been proposed to better trace the convergence behavior of the iterative decoding schemes and to efficiently estimate the thresholds of different codes. However, it is known not to be applicable to the protograph codes \cite{1291808}. In \cite{4411526}, a protograph EXIT (PEXIT) algorithm has been introduced to facilitate the analysis and design of protograph codes over the AWGN channel.
In the following, we illustrate that the PEXIT algorithm, which works well on the AWGN channel, is no longer applicable to a SIMO Rayleigh fading channel. Then we modify the PEXIT algorithm for such a channel and use it for analyzing the protograph codes in our system.

\subsection{{\color{red}Assumption of the PEXIT Algorithm}}
\label{sect:premisePEXIT}
{\color{red}One important assumption of the proposed PEXIT algorithm in \cite{4411526} is that the channel log-likelihood-ratio (LLR) messages should follow a symmetric Gaussian distribution.} In the following, we briefly illustrate that {\color{red} this assumption} cannot be maintained in the case of a SIMO Rayleigh fading channel and then we elaborate how to apply the PEXIT algorithm in such an environment. To  simplify the analysis, we assume that the all-zero codeword is transmitted.

By using $j~(j=1,2,\ldots)$ to indicate the coded bit number and
$k~(k=1,2,\ldots,N_{\rm R})$ to indicate the receive antenna number,
the signal of the $j^{\rm th}$ coded bit at the $k^{\rm th}$ receive antenna can be written as
\begin{equation}\label{eq:receiver_per_antenna}
r_j [k] = h_j [k] x_j + n_j [k]. 
\end{equation}
The combiner output corresponding to the $j^{\rm th}$ coded bit, denoted by $y_j$, is then given by \cite{875282,871398}
\begin{equation}\label{eq:Combiner}
y_j =\left\{\begin{array}{ll}
\displaystyle \sum_{k=1}^{N_{\rm R}} h_j^\ast [k] r_j [k] & ~~\mbox{for MRC} \vspace{0.2mm} \\
\displaystyle \sum_{k=1}^{N_{\rm R}} \frac{h_j^\ast [k]} {|h_j [k]|} r_j [k] & ~~\mbox{for EGC}
\end{array}\right.
\end{equation}
where $^\ast$ denotes the complex conjugate, $|\cdot|$ represents the modulus operator and $h_j^\ast [k] /| h_j [k]|$ is used to remove the phase ambiguity for coherent reception in EGC.
Subsequently, the initial channel LLR value $L_{ch,j}$
corresponding to the $j^{\rm th}$ coded bit 
can be obtained using \cite{4109664} 
\begin{align}
L_{ch,j} &= \ln \left( \frac{ \Pr(v_j=0|y_j,\bh_j)}{ \Pr(v_j=1|y_j,\bh_j)} \right)
= \ln \left( \frac{ \Pr(x_j=+1|y_j,\bh_j)}{ \Pr(x_j=-1|y_j,\bh_j)} \right) \notag\\
&= \left\{\begin{array}{ll}
\displaystyle \frac{2 y_j}{ \sigma_n^2}  & ~~\mbox{for MRC}\\
\displaystyle \frac{2 y_j}{{N_{\rm R}} \sigma_n^2}  \left( \sum_{k=1}^{N_{\rm R}} \left| h_j [k] \right| \right) & ~~\mbox{for EGC}
\end{array} \right.\notag\\
&=\left\{\begin{array}{ll}
\displaystyle \frac{2}{ \sigma_n^2} \left( \sum_{k=1}^{N_{\rm R}} h_j^\ast [k] r_j [k] \right) & ~~\mbox{for MRC}\\
\displaystyle \frac{2}{{N_{\rm R}} \sigma_n^2} \left( \sum_{k=1}^{N_{\rm R}} \frac{h_j^\ast [k]} {|h_j [k]|} r_j [k] \right) \left( \sum_{k=1}^{N_{\rm R}} \left| h_j [k] \right| \right) & ~~\mbox{for EGC}
\end{array} \right.\label{eq:L_oj}
\end{align}
where $\Pr(\cdot)$ is the probability function {\color{red}and $\bh_j=[h_j[1],h_j[2], \cdots,h_j[N_R]]^T $ (the superscript ``$T$'' represents the transpose operator).}

To evaluate the performance of the two combiners, we examine the distribution of $L_{ch,j}$ by exploiting Monte Carlo simulations.
We use a rate-$1/2$ AR3A code with an information length per code block of $1024$. Moreover, we consider a SIMO Rayleigh fading channel with $N_{\rm R} = 2$ and $E_b/N_0 = 2.6$~dB.
By sending $x_j = +1$ repeatedly while varying the channel fading vector $\bh_j$ from bit to bit, we evaluate the mean of the absolute value of $L_{ch,j}$. We
observe that the MRC produces an average
value of $3.628$, i.e., ${\mathbb E}_{\rm MRC} (|L_{ch,j}|)=3.628$ whereas the EGC gives  ${\mathbb E}_{\rm EGC} (|L_{ch,j}|) =3.242$. Moreover, both combiners produce channel LLR values with almost the same variance\footnote{The variance of a complex variable $z$ is given by
          ${\rm var}[z]= {\mathbb E}[(z- {\mathbb E}(z))(z- {\mathbb E}(z))^*]$.}.
Consequently, using MRC should provide a higher chance for successful decoding.
For this reason, \textit{we will focus on MRC in the sequel.}
%
%

\begin{figure}[t]
\center
\includegraphics[width=3in,height=2.5in]{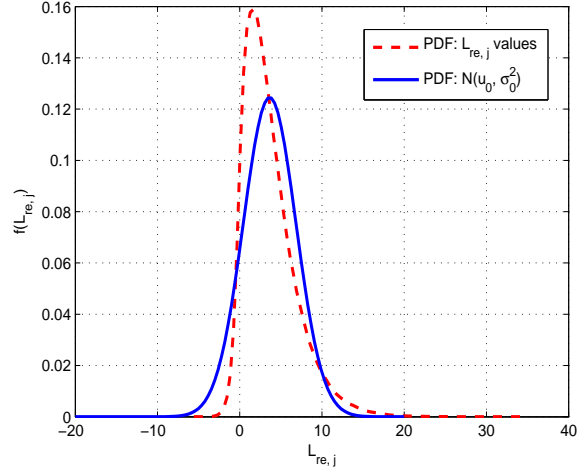}
\vspace{-0.3cm}
\caption{Probability density functions of the $L_{re,j}$ values and $N(u_0, \sigma_0^2)$ over the SIMO Rayleigh fading channel.
}
\label{fig:Fig.2}
\end{figure}

We denote the real part of $L_{ch,j}$ by $L_{re,j}$, i.e., $ L_{re,j}= \Re[L_{ch,j}]$.
In Fig.~\ref{fig:Fig.2},
 we further plot the probability density function (PDF) of the $L_{re,j}$ values (denoted by $f(L_{re,j})$) when MRC is used. We also define $u_0={\mathbb E}_{\rm MRC}(L_{re,j})$ and $\sigma_0^2={\rm var}(L_{re,j})$
 and plot the PDF of the Gaussian distribution ${\cal N}(u_0, \sigma_0^2)$ in the same figure for comparison. The curves in the figure indicate that the PDF of the $L_{re,j}$ values does not agree with the PDF of ${\cal N}(u_0, \sigma_0^2)$, which further suggests that the $L_{ch,j}$ values do not follow a symmetric complex Gaussian distribution\footnote{The same observation is found when EGC is used.}. We conclude that the PEXIT algorithm in \cite{4411526} is not applicable to this type of channel.
In the following, we analyze the distribution of the $L_{ch,j}$ values when the  channel realization is fixed.

We consider a fixed channel realization, i.e., a fixed channel fading vector $\bh_j$.
We assume using the all-zero codeword (i.e., $x_j = +1$) and we substitute \eqref{eq:receiver_per_antenna} 
into \eqref{eq:L_oj}.
Then, we can rewrite the expression for $L_{ch,j}$ as
\begin{align}
L_{ch,j} &=  \frac{2}{ \sigma_n^2} \sum_{k=1}^{N_{\rm R}} h_j^\ast [k] \left( h_j [k] x_j + n_j [k] \right) \notag\\
&=  \frac{2}{ \sigma_n^2} \sum_{k=1}^{N_{\rm R}} \left( \left| h_j [k]\right|^2 + h_j^\ast [k] n_j [k] \right).\label{eq:L_oj_extention}
\end{align}
As ${\mathbb E}\left(n_j [k] n_j^\ast [k]\right) = \sigma_n^2$, we have
\begin{equation}
\left| h_j [k]\right|^2 + h_j^\ast [k] n_j [k] \sim {\cal CN} \left( \left| h_j [k]\right|^2, \left| h_j [k]\right|^2 \sigma_n^2 \right)
\end{equation}
and hence 
\begin{equation}\label{eq:L-PDF}
L_{ch,j} \sim {\cal CN} \left(\frac{2}{\sigma_n^2} \sum_{k=1}^{N_{\rm R}} \left| h_j [k]\right|^2, \frac{4}{\sigma_n^2} \sum_{k=1}^{N_{\rm R}} \left| h_j [k]\right|^2 \right) = {\cal CN} \left(\frac{2}{\sigma_n^2} \alpha_j, \frac{4}{\sigma_n^2} \alpha_j \right)
\end{equation}
where
\begin{equation}
\alpha_j=\sum_{k=1}^{N_{\rm R}} \left| h_j [k]\right|^2
\label{eq:alpha}
\end{equation}
is defined as the \textit{channel factor}.
In this case, we find that the channel LLR values follow a symmetric complex Gaussian distribution.

Besides, assuming $E_s$ is normalized to $1$, \eqref{eq:SNR} can be simplified to $E_b/N_0 = (N_R/R)/N_0$. Thus, combining it with $N_0 = 2 \sigma_n^2$ gives
\begin{equation}\label{eq:sigma_square}
\sigma_n^2 = \frac {N_{\rm R}} {2 R \left( E_b / N_0 \right)}.
\end{equation}

To summarize, we observe that the channel LLR values for a SIMO fading channel do not follow a symmetric Gaussian distribution and hence the PEXIT algorithm cannot be applied directly. However, for a fixed fading vector, these LLR values follow a symmetric complex Gaussian distribution. Using this property, we propose a modified PEXIT algorithm that can be adopted to analyze the protograph codes. Details of the algorithm are described as follows.

\subsection{Modified PEXIT Algorithm for Protograph Codes}
\label{sect:modified PEXIT}
We first define some symbols and terms.
A protograph $G = (V, C, E)$ consists of three sets $V$, $C$ and $E$  corresponding
 to the variable nodes, check nodes and edges, respectively \cite{Thorpe2003ldp}. In a protograph, each edge $e_{i,j} \in E$ connects a variable node $v_j \in V$ to a check node $c_i \in C$. Moreover, parallel edges are allowed.
  A large protograph (namely a derived graph) corresponding to the protograph code can be obtained by a ``copy-and-permute'' operation. Hence, codes with different block lengths can be generated by performing the ``copy-and-permute'' operations different number of times.
  A protograph with $N$ variable nodes and $M$ check nodes can be represented by a base matrix $\bB$ of dimension $M \times N$. The $(i,j)^{\rm th}$ element of $\bB$, denoted by $b_{i,j}$, represents the number of edges connecting the variable node
$v_j$ to the check node $c_i$.
We also define five types of mutual information (MI) as follows.
\begin{enumerate}
\item $I_{Av}(i, j)$ denotes the \textit{a priori} MI between the input LLR value of $v_j$ on each of the  $b_{i,j}$ edges and the corresponding coded bit $v_j$.

\item $I_{Ac}(i, j)$ denotes the \textit{a priori} MI between the input LLR  value of $c_i$ on each of the $b_{i,j}$ edges and the corresponding coded bit $v_j$.

\item $I_{Ev}(i,j)$ denotes the extrinsic MI between the LLR  value sent by $v_j$ to $c_i$ and the corresponding coded bit $v_j$.

\item $I_{Ec}(i, j)$ denotes the extrinsic MI between the LLR  value sent by $c_i$ to $v_j$ and the corresponding coded bit $v_j$.

\item $I_{app}(j)$ denotes the \textit{a posteriori} MI between the \textit{a posteriori} LLR value of $v_j$ and the corresponding coded bit $v_j$.


\end{enumerate}
In addition, during each iteration in the PEXIT algorithm, we have $I_{Ac}(i, j)= I_{Ev}(i, j)$
and $I_{Av}(i, j)= I_{Ec}(i, j)$. We also denote the maximum number of iterations in the algorithm
by $T_{\rm max}$.
Besides, we define two new terms called {\color{red} \textit{indicator function} and  \textit{punctured label}.}

\begin{defi}
We define the {\color{red}\textit{indicator function}} $\Phi(\cdot)$ of an element $b_{i,j}$ in the base matrix $\bB$ as
\begin{equation}
\Phi(b_{i,j}) =
\left\{
\begin{array}{ll}
1 &\;\; \;\;  \text{if} \;\; b_{i,j} \neq 0,  \\
0 & \;\;\;\;   \text{otherwise}.
\end{array}
\right.
\label{eq:T fuction}
\end{equation}
\end{defi}
\noindent Hence, $\Phi(b_{i,j})$ indicates whether $v_j$ is  connected to $c_i$ or not.
\begin{defi}
We define the {\color{red}\textit{punctured label}} $P_j$ of a variable node $v_j$ as $0$ if $v_j$ is punctured, and $1$ otherwise.
\end{defi}

Moreover, the MI between a coded bit and its corresponding LLR value $L_{ch} \sim {\cal N} (\frac{\sigma_{ch}^2}{2}, \sigma_{ch}^2)$
is denoted by
$J(\sigma_{ch})$ and is expressed as \cite{1291808}
  \begin{equation}
J(\sigma_{ch}) = 1- \int_{-\infty}^{\infty}  \frac{\exp \left( -  \frac{(\xi -\sigma_{ch}^2/2)^2 }{2 \sigma_{ch}^2} \right)}{\sqrt{2\pi \sigma_{ch}^2}}
\log_2 \left[ 1+\exp(-\xi) \right] {\rm d} \xi .
\label{eq:J_fuction}
\end{equation}
The inverse function of \eqref{eq:J_fuction} is further given by \cite{1291808}
\begin{equation}
J^{-1}(x) =
\left\{
\begin{array}{ll}
 \gamma_1 x^2 + \gamma_2 x + \gamma_3 \sqrt{x} & \;\; \;\;  \text{if} \;\; 0 \le x \le 0.3646,  \\
\gamma_4 \ln[\gamma_5(1-x)] + \gamma_6 x & \;\;\;\;   \text{otherwise},
\end{array}
\right.
\label{eq:Inverse_J_fuction}
\end{equation}
where $\gamma_1=1.09542, \gamma_2=0.214217, \gamma_3=2.33737, \gamma_4= -0.706692, \gamma_5=0.386013$ and $\gamma_6=1.75017$.

Then, for a rate-$R$ protograph with $N$ variable nodes and $M$ check nodes, the proposed {\bf modified PEXIT algorithm} over a SIMO Rayleigh fading channel can be described as follows.
\begin{enumerate}
\item 
For a given SIMO channel realization $\bh=[h[1],h[2], \cdots,h[N_R]]^T$,  we can calculate the corresponding \textit{channel factor} $\alpha$ using \eqref{eq:alpha}, i.e.,
$\alpha=\sum_{k=1}^{N_{\rm R}} \left| h [k]\right|^2$.
Suppose we are given the number of blocks of channel factors
(denoted by $Q$)
and the maximum number of iterations ($T_{\rm max}$). We generate a matrix $ \bm \alpha = (\alpha_{q,j})
= (\sum_{k=1}^{N_{\rm R}} \left| h_{q,j} [k]\right|^2)$
of dimension $Q \times N$ 
to represent the $Q$ blocks of channel factors,
i.e., each row in
$\bm \alpha$ represents a group of channel factors
for the $N$ variable nodes in the protograph. 
We also select an initial $E_b/ N_0$ (in~{\rm dB})
which should be sufficiently small.

\item \label{it:I_ch}
For $i=1, 2, \ldots, M$ and $j=1, 2, \ldots, N$, we set the initial $I_{Av}(i, j)$ to $0$. We also reset the iteration number $t$ to $0$.
Considering the {\color{red}punctured label} and substituting \eqref{eq:sigma_square} into \eqref{eq:L-PDF}, for the channel factor $\alpha_{q,j}$ ($j=1, 2, \ldots, N$ and $q=1, 2, \ldots, Q$), the corresponding variance of the initial LLR value (denoted by $\sigma_{ch,q,j}^2$) is given by
\begin{equation}\label{eq:sigma_function}
\sigma_{ch,q,j}^2 = \frac{4 P_j \alpha_{q,j}}{\sigma_n^2}
= \frac{8 R P_j \alpha_{q,j}}{N_{\rm R}} 10^{\frac{(E_b/ N_0)}{10}}.
\end{equation}

\item \label{it:I_ev}
If $t= T_{\rm max}$, set $E_b/ N_0=E_b/ N_0 + 0.001$~dB and go to Step \ref{it:I_ch}; otherwise, for  $i=1, 2, \ldots, M$; $j=1, 2, \ldots, N$ and $q=1, 2, \ldots, Q$, we calculate output extrinsic MI sent by $v_j$ to $c_i$ for the $q^{\rm th}$ fading block using \cite{4411526}
\begin{equation}
I_{Ev,q}(i,j) = \Phi(b_{i,j}) J\left(\sqrt{ \left(\sum_{s\neq i} b_{s,j} [J^{-1} (I_{Av}(s,j))]^2 \right) + (b_{i,j} - 1) [J^{-1} (I_{Av}(i,j))]^2 +\sigma_{ch,q,j}^2} \; \right)
\label{eq:I_ev}
\end{equation}

\item
For $i=1, 2, \ldots, M$ and $j=1, 2, \ldots, N$, we obtain the expected value of $I_{Ev,q}(i,j)$ using
\begin{equation}
{\mathbb E} [I_{Ev,q}(i,j)] = \frac{1}{Q} \sum_{q=1}^{Q} I_{Ev,q}(i,j).
\label{eq:AverageI_ev}
\end{equation}
Then, 
the \textit{a priori} MI between the input LLR of $c_i$ on each of the  $b_{i,j}$ edges and the corresponding coded bit is evaluated using
\begin{equation}\label{eq:Evq}
I_{Ac}(i,j)  = {\mathbb E} [I_{Ev,q}(i,j)] .
\end{equation}

\item
For  $i=1, 2, \ldots, M$ and $j=1, 2, \ldots, N$, we compute the output extrinsic MI
  sent by $c_i$ to $v_j$ using \cite{4411526}
\begin{equation}
I_{Ec}(i,j) = \Phi (b_{i,j}) \left(1- J\left(\sqrt{\left(\sum_{s\neq j} b_{i,s} [J^{-1} (1 - I_{Ac}(i,s))]^2 \right) + (b_{i,j} - 1) [J^{-1} (1 - I_{Ac}(i,j))]^2}\right) \right)
\label{eq:I_ec}
\end{equation}
Then, 
we get the \textit{a priori} MI between the input LLR of $v_j$ on each of the  $b_{i,j}$ edges and the corresponding coded bit using
\begin{equation}
I_{Av}(i,j) = I_{Ec}(i,j) .
\label{eq:AverageI_ec}
\end{equation}

\item
For $j=1, 2, \ldots, N$ and $q=1, 2, \ldots, Q$, we compute the \textit{a posteriori} MI of $v_j$ using \cite{4411526}
\begin{equation}
I_{app,q}(j) = J \left(\sqrt{\left(\sum_{s=1}^{N} b_{s,j} [J^{-1} (I_{Av}(s,j))]^2 \right) +\sigma_{ch,q,j}^2}\;\right) .
\label{eq:I_app}
\end{equation}
Then, for every $j=1, 2, \ldots, N$, we can evaluate the expected value of $I_{app,q}(j)$ using
\begin{equation}
{\mathbb E}[I_{app,q}(j)] = \frac{1}{Q} \sum_{q=1}^{Q} I_{app,q}(j) .
\label{eq:AverageI_app}
\end{equation}

\item
If the expected MI values ${\mathbb E}[I_{app,q}(j)] = 1$ for all $j = 1, 2, \ldots, N$,
the $E_b/N_0$ value will be the EXIT threshold that allows all variable nodes to be decoded correctly and the iterative process is stopped; otherwise, we increase $t$ by $1$ and go to Step \ref{it:I_ev} to continue the iterative process.
\end{enumerate}

Note also the following.
\begin{itemize}
{\color{red}
\item Each of the output extrinsic MI sent by $v_j$ to $c_i$ depends on both the initial channel LLR values $\sigma_{ch,q,j}^2$ \eqref{eq:sigma_function} AND the type of code used.
Thus, for a fixed $(i,j)$, we can compute $Q$ different values for $I_{Ev,q}(i, j)$ \eqref{eq:I_ev}.
Consequently, an average quantity ${\mathbb E} [I_{Ev,q}(i,j)]$ \eqref{eq:AverageI_ev} can be obtained based on the $Q$ different values of $I_{Ev,q}(i, j)$.

\item The average quantity ${\mathbb E} [I_{Ev,q}(i,j)]$ becomes the  \textit{a priori} MI $I_{Ac}(i,j)$ between the input LLR of $c_i$ on each of the  $b_{i,j}$ edges and the corresponding coded bit \eqref{eq:Evq}.

\item Since the output extrinsic MI sent by $c_i$ to $v_j$ is only related to
the type of code used and is independent of the initial channel LLR values $\sigma_{ch,q,j}^2$,
 the output extrinsic MI sent by $c_i$ to $v_j$, i.e., $I_{Ec}(i,j)$, can be computed using
\eqref{eq:I_ec}, which depends only on the   \textit{a priori} MI $I_{Ac}(i,j)$  ($i=1, 2, \ldots, M$ and $j=1, 2, \ldots, N$) and is independent of $\sigma_{ch,q,j}^2$.
For a fixed $(i,j)$, there is only one  $I_{Ec}(i,j)$ computed and thus no averaging is required.

\item The quantity $I_{Ec}(i,j)$ then becomes the \textit{a priori} MI $I_{Av}(i,j)$ between the input LLR of $v_j$ on each of the  $b_{i,j}$ edges and the corresponding coded bit  \eqref{eq:AverageI_ec}.

\item Each of the \textit{a posteriori} MI value  $I_{app,q}(j)$ of $v_j$ depends on both the initial channel LLR values $\sigma_{ch,q,j}^2$ AND the type of code used.
Thus, for a fixed $j$, we can compute $Q$ different values for $I_{app,q}(j)$  \eqref{eq:I_app}.
Afterwards, an average quantity ${\mathbb E}[I_{app,q}(j)]$  \eqref{eq:AverageI_app} can be obtained based on the $Q$ different values of $I_{app,q}(j)$.
}
%

\item
To ensure the accuracy of the modified PEXIT algorithm, we should generate a sufficiently large number of blocks of channel factors, i.e., a large value for $Q$. In this paper, we use $Q = 10^5$.
\end{itemize}

\section{Analysis of Protograph Codes}\label{sect:analysis}
Protograph codes  not only enable linear encoding and decoding to be implemented easily, but also have superior error performance over the AWGN channel \cite{Thorpe2003ldp}. As two typical LDPC codes constructed by protographs, the AR3A
code and the AR4JA code  possess excellent performance in the waterfall region and the error floor region, respectively, over the AWGN channel \cite{1577834,4155107,5174517,5454136}. The corresponding base matrices of the AR3A code and the AR4JA code with a code rate $R=(n+1)/(n+2)$ are denoted by $\bB_{A3}$ and $\bB_{A4}$, respectively, where
\begin{align}
\bB_{A3} &=\left(\begin{array}{llllll}
1 & 2 & 1 & 0 & 0  & \overbrace{0 \ 0 \ \cdots \ 0 \ 0}^{2n} \cr
0 & 2 & 1 & 1 & 1  & 2 \ 1 \ \cdots \ 2 \ 1 \cr
0 & 1 & 2 & 1 & 1  & 1 \ 2 \ \cdots \ 1 \ 2
\end{array}\right) \label{eq:BA3}\\
\bB_{A4} &=\left(\begin{array}{llllll}
1 & 2 & 0 & 0 & 0  & \overbrace{0 \ 0 \ \cdots \ 0 \ 0}^{2n} \cr
0 & 3 & 1 & 1 & 1  & 3 \ 1 \ \cdots \ 3 \ 1 \cr
0 & 1 & 2 & 2 & 1  & 1 \ 3 \ \cdots \ 1 \ 3
 \end{array}\right).\label{eq:BA4}
\end{align}
We assume that the $j^{\rm th}$ column of the matrix corresponds to the $j^{\rm th}$ variable node and the $i^{\rm th}$ row of the matrix corresponds to the $i^{\rm th}$ check node. Note that the variable nodes corresponding to the second columns in \eqref{eq:BA3} and \eqref{eq:BA4} are punctured.

Using the modified PEXIT algorithm proposed in Sect.~\ref{sect:modified PEXIT}, we firstly investigate the decoding thresholds of the AR3A and AR4JA codes with a code rate of $R = 1/2$ (i.e., when $n = 0)$. For comparison, the regular $(3, 6)$ LDPC code and the two optimized irregular LDPC codes (denoted as irregular LDPC code A and irregular LDPC code B) in \cite{4109664} are used to gauge the performance. Moreover, the degree distribution pairs of the irregular codes are given as
\begin{equation}\label{eq:A-distribution}
\left\{\begin{aligned}
\lambda_{\rm A}(x)&=0.270234x + 0.266315x^2 + 0.463451x^9\\
\rho_{\rm A}(x)&= 0.566545x^6 + 0.433455x^7
\end{aligned}\right.
\end{equation}
and
\begin{equation}\label{eq:B-distribution}
\left\{\begin{aligned}
\lambda_{\rm B}(x)&=0.285637x + 0.285602x^2 + 0.428724x^8\\
\rho_{\rm B}(x)&= 0.998857x^6 + 0.001143x^7
\end{aligned}\right.
\end{equation}
As shown in \cite{4109664}, irregular code A and irregular code B are optimized for the diversity orders of $N_{\rm R} =2$ and $N_{\rm R} =4$,
respectively. For our system, the diversity order equals $N_{\rm R}$. Table~\ref{tab:Thre-Compare} shows the decoding thresholds and
the \textit{capacity gaps} $\triangle$\footnote{The capacity gap is defined as the distance between the channel capacity and the decoding threshold.} of these four codes over SIMO Rayleigh fading channels with two different diversity orders. Results in this table indicate that
the decoding thresholds of irregular code A and irregular code B
are smallest over such channels with the diversity
order $2$ and $4$, respectively. These small thresholds suggest that the irregular codes should possess relatively better performance in the
low-SNR region
\cite{4155107,5174517}. However, the irregular codes may be outperformed by other codes in the high-SNR region because of the error-floor issue. Also, the results demonstrate that the regular
LDPC code possesses the highest threshold. Hence, the regular code is expected to be inferior in the low-SNR region. Between the AR3A code
and the AR4JA code, it is further observed that the AR3A code has lower thresholds for both the cases of $N_{\rm R} = 2$ and $N_{\rm R} = 4$.
\begin{table*}[t]
\caption{Decoding thresholds $(E_b/N_0)_{\rm th}$~({\rm dB}) and the capacity gaps $\triangle$ of the AR3A code, AR4JA code, regular $(3, 6)$ code, and the optimized irregular LDPC codes (irregular code A for $N_{\rm R} = 2$ and irregular code B for $N_{\rm R} = 4$) over the SIMO Rayleigh fading channels. The parameters used are $n = 0$ and $R = 0.5$.}
\centering
\begin{tabular}{|c|c|c|c|c|c|c|c|c|c|}
    \hline
    \multirow{2}{*}{$N_{\rm R}$} & \multirow{2}{*}{Capacity}
    & \multicolumn{2}{|c|}{AR3A code} & \multicolumn{2}{|c|}{AR4JA code} & \multicolumn{2}{|c|}{Regular code} & \multicolumn{2}{|c|}{Irregular code in \cite{4109664}}\\
    \cline{3-10}
    & & {${(E_b/N_0)}_{\rm th}$} & $\triangle$ & {${(E_b/N_0)}_{\rm th}$} & $\triangle$ & {${(E_b/N_0)}_{\rm th}$} & $\triangle$ & {${(E_b/N_0)}_{\rm th}$} & $\triangle$\\
    \hline
    $2$  & $-0.514$ & $1.258$ & $1.772$	& $1.433$ & $1.947$ & $1.993$ & $2.507$ &	 $1.175$	& $1.689$ \\
    \hline
    $4$	& $-0.662$ & $0.871$ &	$1.533$ & $1.011$ &	$1.673$	& $1.535$ & $2.197$	& $0.823$	& $1.485$ \\
    \hline
\end{tabular}
\label{tab:Thre-Compare}
\vspace{-0.2cm}
\end{table*}
\begin{table*}[t]
\caption{Decoding thresholds ${(E_b/N_0)}_{\rm th}$~(in {\rm dB}) of the AR3A code and AR4JA code with different code rates $R$ over the SIMO Rayleigh fading channels with diversity orders $N_{\rm R} = 1$, $2$, $3$, and $4$.}
\centering
\begin{tabular}{|c|c|c|c|c|c|c|c|c|}
    \hline
    \multirow{2}{*}{Code Rate}
    & \multicolumn{4}{|c|}{AR3A code} & \multicolumn{4}{|c|}{AR4JA code}\\
    \cline{2-9}
    & $N_{\rm R} = 1$ &$N_{\rm R} = 2$ & $N_{\rm R} = 3$ & $N_{\rm R} = 4$
    & $N_{\rm R} = 1$ &$N_{\rm R} = 2$ & $N_{\rm R} = 3$ & $N_{\rm R} = 4$ \\
    \hline
    $1/2 \, (n=0)$ & $2.143$ &	$1.258$ & $1.031$ &	$0.871$ & $2.303$ &	$1.433$ & $1.151$ & $1.011$ \\
    \hline
    $2/3 \, (n =1)$ & $3.898$ & $2.540$ & $2.105$ & $1.903$ & $4.134$ & $2.752$ &
    $2.293$ & $2.084$ \\
    \hline
    $3/4 \, (n =2)$ & $5.154$ & $3.402$ & $2.847$ & $2.609$ & $5.413$ & $3.624$ &
    $3.068	$ & $2.783$\\
    \hline
    $4/5 \, (n =3)$ & $6.175$ & $4.137$ & $3.479$ & $3.156$ & $6.506$ & $4.322$ &
    $3.648$ & $3.315$\\
    \hline
    $5/6 \, (n =4)$ & $6.985$ & $4.698$ & $3.960$ & $3.616$ & $7.286$ & $4.871$ &
    $4.117$ & $3.746$\\
    \hline
    $6/7 \, (n =5)$ & $7.667$ & $5.129$ & $4.361$ & $3.969$ & $7.996$ & $5.294$ &
    $4.490$ & $4.103$\\
    \hline
    $7/8 \, (n =6)$ & $8.399$ & $5.559$ & $4.691$ & $4.292$ & $8.644$ & $5.691$ &
    $4.830$ & $4.406$\\
    \hline
\end{tabular}
\label{tab:Thre-diff-orders}
\vspace{-0.2cm}
\end{table*}

We further investigate the performance of the AR3A code and the AR4JA code in such systems with different code rates and different diversity orders. The results are shown in Table~\ref{tab:Thre-diff-orders}. Given a fixed code rate, we observe from the table that the decoding threshold of the AR3A code is smaller than that of the AR4JA code for all diversity orders under study.
We also find that   as the diversity $N_{\rm R}$ increases, the threshold decreases and hence the error performance of both codes should improve in the low-SNR region. The decrease in the threshold, however, is reduced as
$N_{\rm R}$ is incremented. For example, we consider the AR3A code with a rate of $R=1/2$. The threshold reduces from $2.143$ to $1.258$
as $N_{\rm R}$ increases from $1$ to $2$; but further reduces to
 $1.031$ and  $0.871$ only when $N_{\rm R}$ increases to $3$ and $4$, respectively.
%

For the AR3A code, we also evaluate the mean and the variance of the initial LLR values output by the MRC combiner using Monte Carlo simulations. The results, which are generated with the same parameters used for creating Fig.~\ref{fig:Fig.2}, are given in Table~\ref{tab:mean-var}.
The results show that both (i) the rate that the mean of $|L_{ch,j}|$ increases and (ii)  the rate that the variance of $L_{ch,j}$ decreases are reduced as
 $N_{\rm R}$ is incremented. Similar simulations have
further been performed for the AR4JA code and the same observations are obtained. These results imply that the rate of improvement of the error performance reduces when we increase the diversity order.
\begin{table}[t]
\caption{The mean and variance of the initial LLR values output by the MRC combiner over the SIMO Rayleigh fading channels. An AR3A code with $n=0$ and $R=1/2$ is used.  The information length per code block is $1024$ and $E_b/N_0 = 2.6$~{\rm dB}. Diversity order $N_R=1$, $2$, $3$, and $4$.}
\begin{center}
\begin{tabular}{|c|c|c|c|c|}
\hline
$N_{\rm R}$ & $1$ &  $2$  & $3$  & $4$ \\\hline
${\mathbb E} (|L_{ch,j}|)$ & $3.615$ & $3.628$ & $3.634$ & $3.636$ \\\hline
${\rm var} (L_{ch,j})$ & $20.251$ & $13.053$ & $11.669$ & $10.502$ \\\hline
\end{tabular}
\end{center}
\label{tab:mean-var}
\end{table}%

\section{Simulation Results}\label{sect:sim_dis}
In this section, we simulate the error performance of the AR3A code, the AR4JA code, the regular $(3,6)$ code, and the optimized irregular LDPC codes in \cite{4109664} over SIMO Rayleigh fading channels. We also discuss the influence of the diversity order on the error performance of the protograph codes. The code rate and the information length of each code block are $R = 1/2$ and $1024$, respectively. We terminate the simulations after $500$ bit errors are found at each $E_b/N_0$. Also, the LDPC decoder performs a maximum of $100$ BP iterations for each code block.

\subsection{Performance Comparison among Four Different Codes}
\begin{figure}[t]
\centering
\subfigure[\hspace{-0.8cm}]{ \label{fig:subfig:a} 
\includegraphics[width=2.8in,height=2.2in]{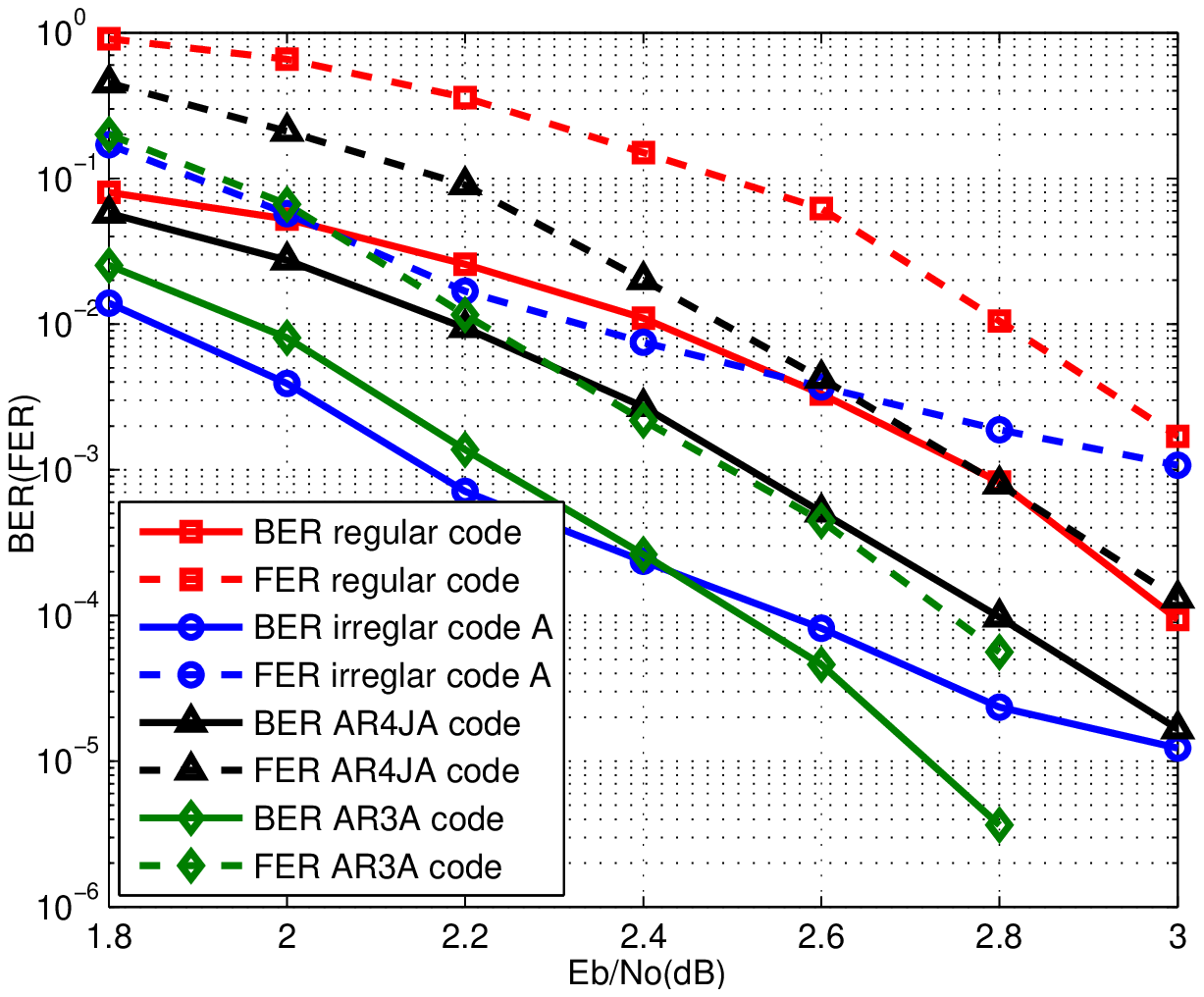}}
\subfigure[\hspace{-0.8cm}]{ \label{fig:subfig:b} 
\includegraphics[width=2.8in,height=2.2in]{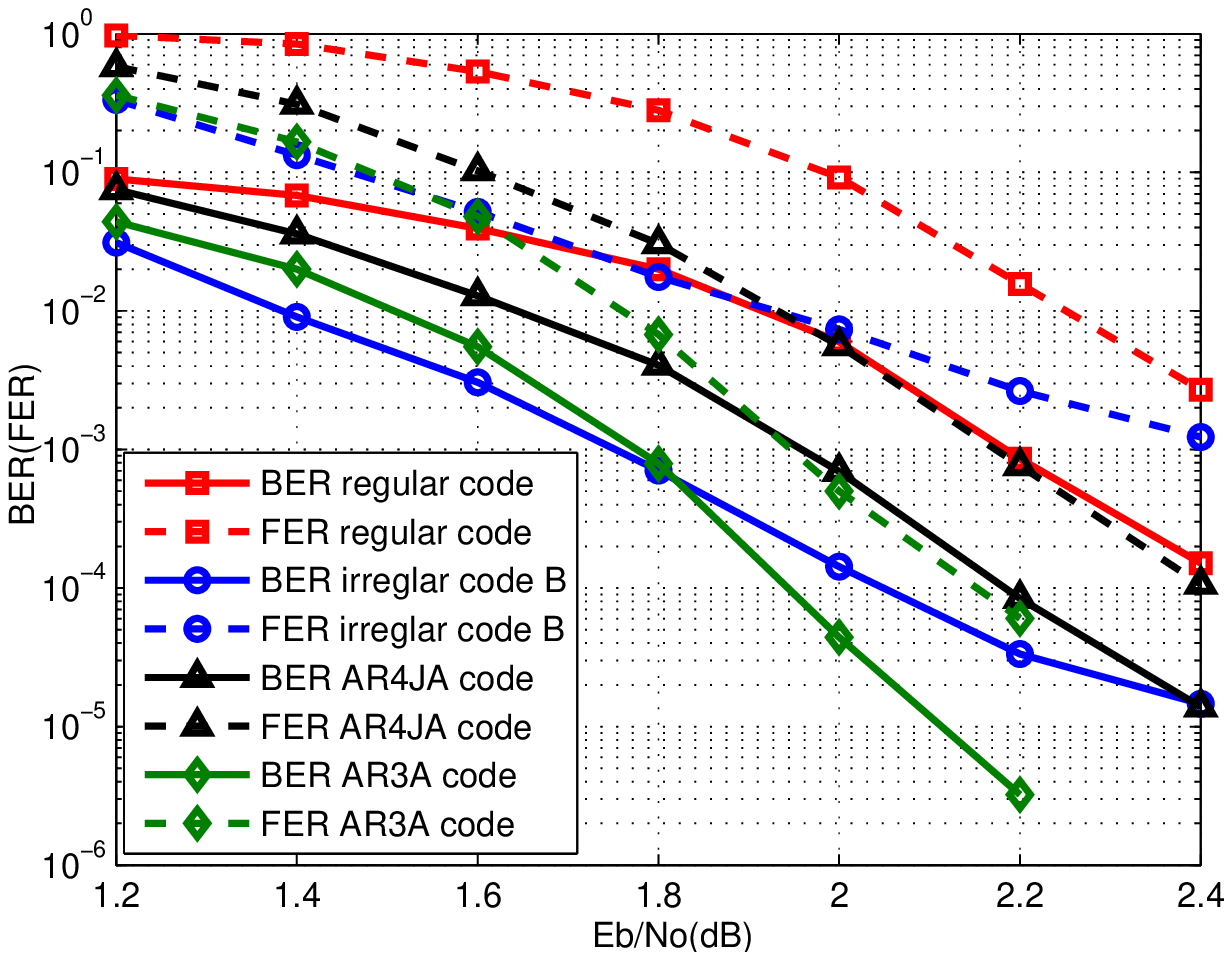}}
\vspace{-0.3cm}
\caption{BER and FER curves of the AR3A code, the AR4JA code, the regular LDPC code, and the irregular LDPC codes over the SIMO Rayleigh fading channels
with the diversity order (a) $N_{\rm R}=2$ and (b) $N_{\rm R}=4$.}
\vspace{-0.1cm}
\label{fig:Fig.3}  
\end{figure}

Figure~\ref{fig:Fig.3} plots the bit error rate (BER) and frame error rate (FER) curves of the codes over the SIMO Rayleigh fading channels with diversity orders $N_{\rm R}=2$ and $N_{\rm R}=4$.
It can be seen that the AR4JA code and the regular $(3, 6)$ code are the two worst-performing codes for the two diversity orders.
Moreover, referring to Fig.~\ref{fig:Fig.3}(a), at a BER of $10^{-3}$, the irregular code A has a gain of about $0.1$~dB over the AR3A code, which remarkably outperforms the other two codes. Yet, the BER and the FER performance of the irregular LDPC codes has little improvement when the $E_b/N_0$ exceeds $2.4$~dB upon which the error floor emerges. In the same figure, we observe that the AR3A code has excellent error performance for the range of $E_b/N_0$ under study. For instance, at $E_b/N_0 = 2.8$~dB, the AR3A code accomplishes a BER of $4 \times 10^{-6}$, while the irregular code,
AR4JA code, and the regular code achieve BERs of $2 \times 10^{-5}$, $10^{-4}$, and $8 \times 10^{-4}$, respectively. We also observe that at a BER of $10^{-5}$, the AR3A code has a performance gain of $0.3$~dB over the irregular LDPC code and the AR4JA code, which are superior to the regular code. Moreover, a large gain
can be expected at a lower BER. Similar conclusions can be drawn from Fig.~\ref{fig:Fig.3}(b), where the diversity order equals $4$.

In general, among the four types of codes, we conclude that the optimized irregular codes possess the best error performance in the low-SNR region while the
AR3A code can provide excellent error performance in the high-SNR region over the SIMO Rayleigh fading channels.

\subsection{Discussion about the Diversity Order}
\begin{figure}[htbp]
\center
\includegraphics[width=2.8in,height=2.2in]{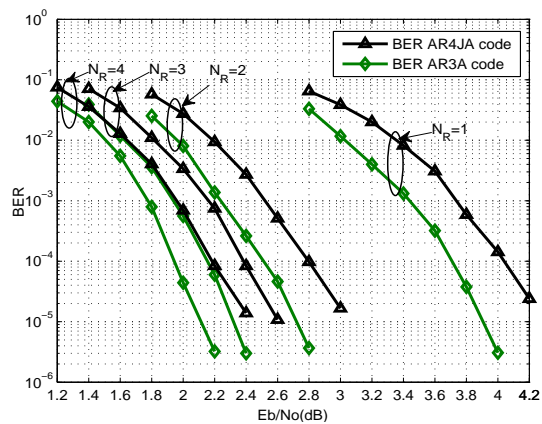}
\caption{The BER results of the AR3A code the AR4JA code over the SIMO Rayleigh fading channels with diversity orders $N_{\rm R}=1$, $2$, $3$, and $4$.}\label{fig:Fig.4}
\end{figure}

Figure~\ref{fig:Fig.4} shows the BER results of the AR3A code and the AR4JA code for various diversity orders ($N_{\rm R}=1$, $2$, $3$, and $4$).
As can be seen from this figure, the AR3A code outperforms the AR4JA code by more than $0.3$~dB at a BER of $2 \times 10^{-5}$
for all diversity orders. We also observe that for a fixed BER, the required $E_b/N_0$ decreases as the diversity order increases. However, the rate of decrease is reduced with the diversity order. For example, we consider the AR4JA code at a BER of $2 \times 10^{-5}$. The required $E_b/N_0$ decreases from $4.2$~dB to $3.0$~dB, $2.5$~dB and $2.3$~dB, respectively, as $N_{\rm R}$ from $1$ to $2,3$ and $4$.
Similar observations are found for the AR3A code.
Furthermore, these observations agree well with the analytical results found in Section~\ref{sect:analysis}. On the other hand, the system complexity  increases with the diversity order.
Accordingly, one should appropriately select the number of the receive antennas for a  practical system so as to make a good balance between system
performance and implementation complexity.

\section{Conclusions}\label{sect:conclusion}
In this paper, we have studied the performance of protograph-based LDPC codes for SIMO systems under fading channel conditions. We have proposed a modified PEXIT algorithm for analyzing such systems equipped an MRC combiner. We have also used the proposed algorithm to evaluate the decoding threshold of the protograph codes and hence to analyze their error performance. Furthermore, we have compared the decoding thresholds and the distribution of the initial LLR values of the protograph codes among different diversity orders. We conclude that
while the error performance of the protograph codes improves with the diversity order, the rate of improvement is reduced as the diversity order becomes larger.
Thus, we have to strike a balance between code performance and system complexity when determining the diversity order to be implemented.
Our simulation results have also shown that the AR3A code is able to provide a significant gain over the AR4JA code, the regular (3, 6) code, and the optimized irregular LDPC codes in the high-SNR region. In the future, we will strive to optimize the protograph codes in such systems. Furthermore, we will explore extending the modified PEXIT algorithm to other systems such as the MIMO fading systems.


\end{document}